\documentclass[a4paper,11pt]{article}
\usepackage{jinstpub} 
\usepackage{lineno}
\linenumbers
\usepackage{subcaption}
\proceeding{
The 16th Topical Seminar on Innovative Particle and Radiation Detectors (IPRD23)\\
25 September to 29 September 2023\\
Siena}

\title{\textbf{Automated object detection for muon tomography data analysis}}

\author[a,b,1]{Anzori Sh. Georgadze,\note{Corresponding author.}} 
\affiliation[a]{Institute of Physics, University of Tartu,\\ W. Ostwaldi 1, 50411, Tartu, Estonia}
\affiliation[b]{ Institute for Nuclear Research of the National Academy of Sciences of Ukraine,\\Prospekt Nauky 47, 03680, Kyiv, Ukraine}
\emailAdd{anzori.heorhadze@ut.ee}

\abstract{
In recent years, there have been ongoing efforts to improve screening technologies to improve security and prevent terrorist threats. The most widely used technologies for scanning shipping containers are gamma and x-ray radiography, which can be harmful to operators and the environment. Muon tomography screening systems are considered as a potential tool to enhance border security and prevent terrorist threats or smuggling, especially in the context of shipping container inspections. Muon tomography is a technique that uses naturally occurring cosmic ray muons to create detailed images of the inside of objects, such as shipping containers, without the need for physical intervention. In this paper, we describe the algorithms that allow automatic detection of illegal dangerous items hidden inside legal cargo in shipping containers. We used the Point of Closest Approach (PoCA) reconstruction algorithm to reconstruct the 3D image of the shipping container and applied the nearest neighbor filtering method to separate and differentiate the filament materials from the complex structure of the surrounding common materials. The use of these methods makes it possible to identify implicit information in the data and visualize the contents of transport containers with precise localization of threats or contraband materials.
}

\keywords{Computerized Tomography (CT) and Computed Radiography (CR), Data processing methods, Image filtering, Detection of contraband and drugs}
\arxivnumber{1234.56789} 
\begin{document}
\nolinenumbers
	\maketitle
	\flushbottom
	\keywords{First keyword \and Second keyword \and More}

\section{Introduction}
The rapid growth in international commerce has presented a significant challenge for cargo security, and one of the primary issues is the limited capacity to inspect and verify the contents of containers. Only a small percentage of containers are typically scanned or inspected thoroughly at ports and border crossings which leaves open the possibility of security breaches and smuggling activities. 
The cargo imaging inspection systems are aimed at finding illegal materials, such as weapons, explosives, drugs or Special Nuclear Materials (SNM) through the imaging of large objects, such as cargo containers, unoccupied vehicles, trains, trucks or boats. 
Traditional methods of cargo inspection, such as X-ray scanning and manual inspections, are often time-consuming and may not provide a comprehensive view of the contents of containers or cargo.  Cosmic-ray tomography is an innovative method for cargo security and inspection that has the potential to address the challenges posed by the rapid growth in international commerce. 
Muons are subatomic particles that are naturally produced when cosmic rays interact with the Earth's atmosphere. They have the unique property of being highly penetrating, which means they can pass through thick materials like concrete and metal. This property makes them suitable for cargo inspection, as they can be used to create detailed images of the interior of containers or cargo without the need for physical intrusion. 
Muons interact with matter primarily through the electromagnetic force which makes muons to deflect from their original trajectory (Coulomb scattering) and causes the reduction in muon energy (mainly via ionisation and excitation energy loss mechanism). 
Cosmic-ray muons are part of the natural background radiation environment that humans are exposed to continuously and hence, will not harm or alter the objects being investigated, making them ideal for non-invasive inspection and imaging. 
Cosmic-ray muons have energies at sea level about 3–4 GeV and are able to penetrate much deeper into the materials compared to X-rays. 
Cosmic-ray tomography~\cite{bonechi,barnes2023cosmic, borozdin2003, explosives, yifan2018discrimination, lowZ} is considered to be a new inspection technology, complementary to the X-ray imaging which is most widely used to scan cargo at borders. It is expected that CRT will help to increase the number and fraction of scanned cargoes. 

This paper investigates the possibility of applying cosmic-ray tomography for scanning shipping containers and lorries for contraband hidden inside legal cargo within acceptable measuring time. We have used GEANT4 toolkit~\cite{geant4} for simulating realistic scenarios of illegal materials hidden inside legal cargo.

\section{GEANT4 simulation and tomographic reconstruction}

Simulation model  of  the  muon tomography system consisting of two tracking modules above and below the shipping container was constructed with the GEANT4 toolkit~\cite{geant4}. 
Each tracking module, having two position-sensitive detectors, was modeled as a plane of plastic scintillators with a detection efficiency of 100\% and an area of 8 m $\times $4 m and fit that of a shipping container.

\begin{figure*}[t]
	\begin{minipage}{1.\linewidth}
		\centering
		\includegraphics[width=0.7\linewidth]{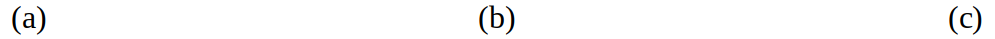}
		\vspace{-1.mm}  
	\end{minipage}		
	\begin{minipage}[t]{0.33\textwidth}
		\centering
		\includegraphics[width=0.99\textwidth]{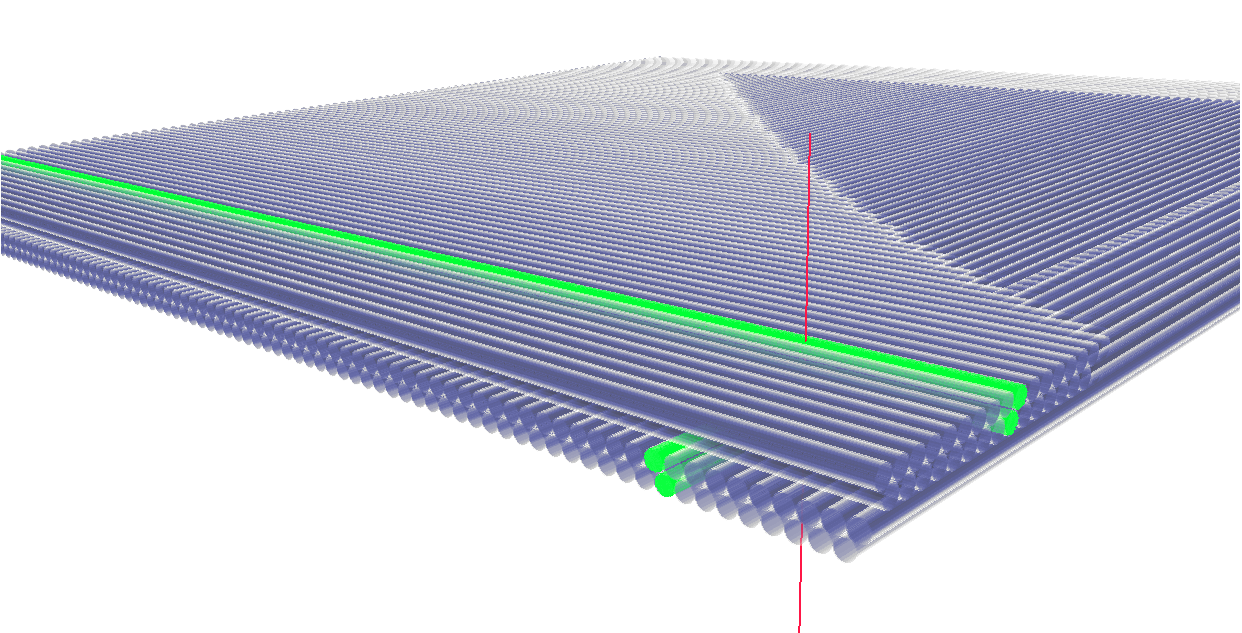}
	\end{minipage}
	\begin{minipage}[t]{0.3\textwidth}
		\centering
		\includegraphics[width=1.0\textwidth]{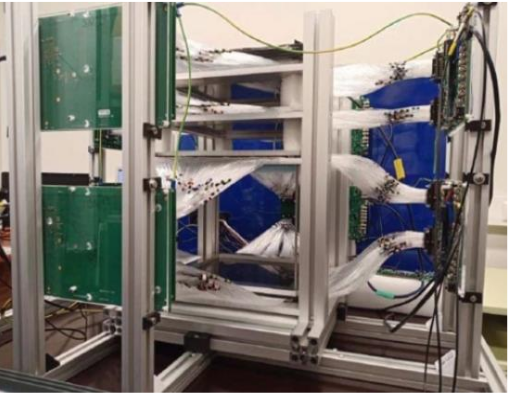}
	\end{minipage}
	\begin{minipage}[t]{0.33\textwidth}
	\centering
	\includegraphics[width=1.0\textwidth]{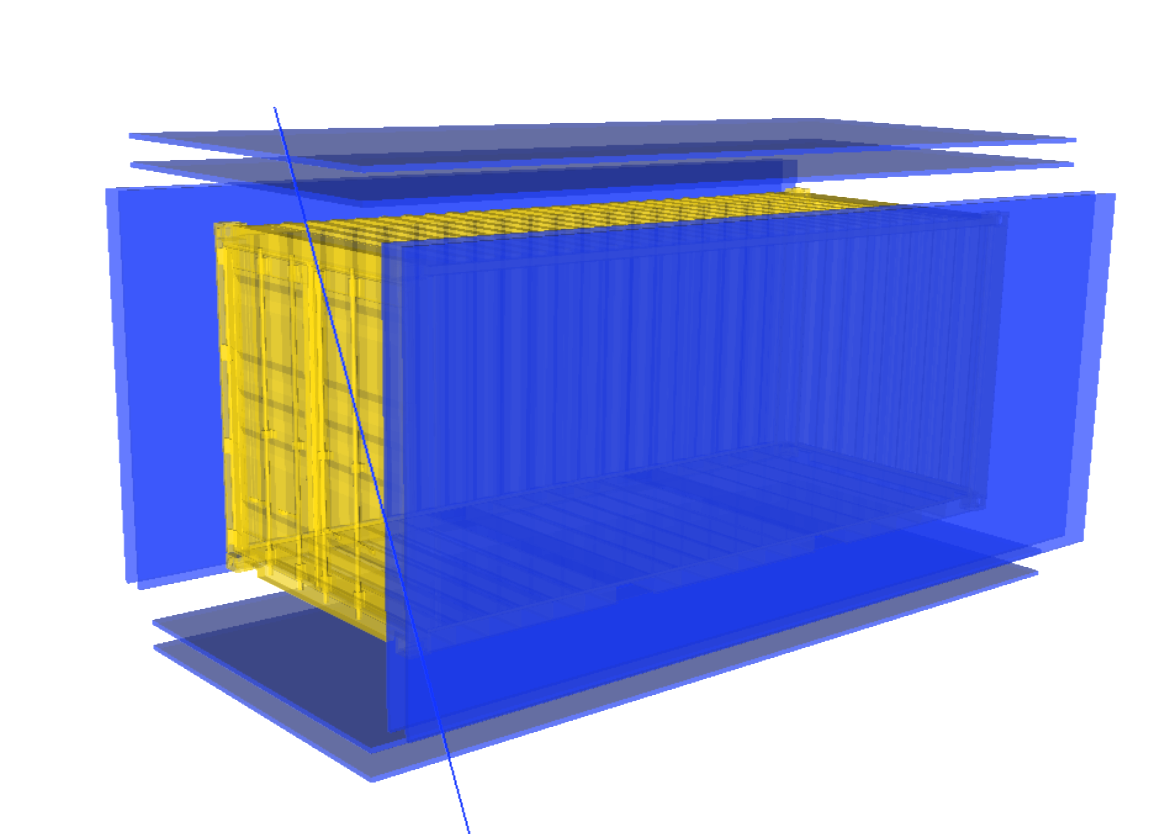}
\end{minipage}
	\caption{(a) Layout of the plastic scintillator fiber tracker in GEANT4 with optical photons  generated in fibers by muon; (b) plastic scintillator fiber tracker with an area around $30\times30$ cm$^2$ constructed in~\cite{Anbarjafari}; (c) illustration cosmic ray tomography (CRT) system.} 
	\label{fig:figure1}      
\end{figure*}

As a possible tracking detector concept we consider technology based on plastic scintillating fiber arrays readout with Silicon Photomultipliers (SiPMs)~\cite{Anbarjafari}. The tracker consists of two planes, with the two layers of fibers in each plane oriented perpendicularly to those in the adjacent plane, which allow three-dimensional reconstruction of the muon track (figure~\ref{fig:figure1}). 
The Cosmic-Ray Shower Library (CRY)~\cite{hagmann2007cosmic} was used to generate muons for the latitude 50° and altitude at sea-level. The origin points of generated muons we sampled from a horizontal plane surface of 10 m $\times$ 10 m. Generated muons are interfaced to GEANT4 to simulate the interaction of the muons with the detector, shipping container and cargo. The simulated data samples were produced with an exposure time ranging from 5 to 10 minutes (1 million muons correspond to ~1 minute of scanning time). The recorded data on hit position on each detector layer are used to produce two tracks. Cosmic ray muon scattering tomography is achieved by tracking the trajectories of both incoming and outgoing muons using a set of position-sensitive detectors.
We calculate scattering angle between the trajectories of incoming ($\vec{v}_1$) and outgoing ($\vec{v}_2$) muons using formula ~\cite{carlisle2012multiple}:
\begin{equation}\label{key}
\theta_{scatt} = arccos\left ( \frac{\vec{}v_{1}\cdot \vec{}v_{2}}{\left | v_{1} \right | \left | v_{2} \right |} \right ). 
\end{equation}
Since some of the hits create statistical noise due to the applied detector resolution, several selection criteria were implemented to select only relevant scattering points for track recovery - The muon tracks must pass through all detectors.
The scattering point must be located inside the RoI. The angle between the tracks constructed from any pair of detectors in the top and bottom set must not exceed 10 mrad. We use the standard physics list for high-energy transport named "FTFP\_BERT". Synthetic datasets were analyzed using ROOT data analysis package ~\cite{ROOT}. 

For the reconstruction of tomographic images we have used Point-of-Closest-Approach algorithm (POCA) ~\cite{PoCA}, which is based on the assumption that muon scattering occurs in a single point. In order to take into account detector position resolution we reconstruct the POCA points with hit positions in the tracking detector smeared simultaneously in x and y directions by Gaussian with the resolution $\sigma$ of 0.15 mm obtained in~\cite{Anbarjafari}. 

\begin{figure}[t]
	\centering
	\includegraphics[width=1.\linewidth]{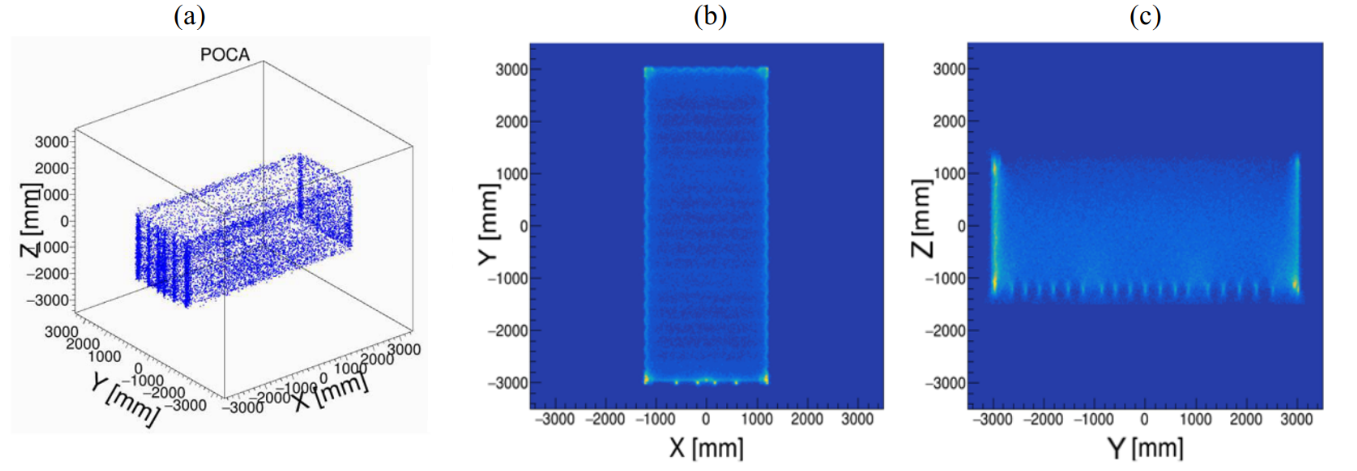} 
	\caption{ Visualization of CRT system with shipping container using GEANT4 (a), XY (b) and YZ (c) projections of tomography image of empty shipping container} 
	\label{fig:figure2}  
\end{figure}

In order to obtain a clearer image with better signal-to-background ratio and determine positions of hidden illegal goods and an estimate of their density besides image filtering we perform spatial cuts, removing PoCA points outside shipping container area. Next we subtract the image of an empty container (figure~\ref{fig:figure2} ) from the 3D image of a loaded container. Strict standards of shipping containers make the procedure of image subtraction quite accurate. After subtraction we apply additional spatial cut of an image to dimensions $5.85\times2.35\times2.35$ m$^{3}$ which correspond to internal  part of the container. 

We apply statistical filtering method to remove noise by finding a median value among voxels in 3D image. Median filtering removes noise by replacing the pixel or voxel value at a given location with the median value of the pixels in a local neighborhood around that location.
This method allow to clear image of areas with lower scattering density comparing comparing to large scattering density generated by most of illegal materials. In some cases, hidden contraband materials has lower density then material matrix of legal cargo. In this case hidden contraband will be visualized on image as a gap or hole and will be recognized as anomaly and case for additional investigation. 
If illegal material has a higher density than legal cargo, it can be detected by the difference in scattering density of materials. 
Since illegal materials are hidden in massive surrounding cargo material we use an algebraic filtering algorithm to remove voxels that represent material of cargo to focus on anomalies that might indicate the presence of illegal materials.
To do that the nearest neighboring voxels are examined. 
The filtering algorithm compares the value of a voxel with a threshold value that has a functional dependence on the average scattering density around the voxel. 
This dependence is a function of the variance of the scattering density around a given voxel in a 3D grid. 
In detail, each voxel has 26 adjacent voxels in the immediate surrounding layer. Moving to the second layer around the given voxel, there would be 80 voxels, not including the original voxel or the immediate neighbors. The applied algorithm evaluates up to ten layers around a given voxel. This multilevel estimation allows the threshold to vary depending on the scattering density in the far environment around the voxel.
This algorithm is particularly effective at removing background from medium-Z materials such as steel, and aggregating voxels containing high-Z materials when searching for smuggled Special Nuclear Materials (SNMs) in cargo.

\section {Simulation in a real scenarios} 
To evaluate the ability of muon tomography to detect illegal materials in a real-world environment, we generated geometries in Geant4 of shipping container loaded with legal cargo and hidden inside illegal goods. 
The sheer volume and complexity of the flow of commercial goods make it impractical to model every possible smuggling scenario. Simulation of a limited number of scenarios serves as the basis for the development of artificial intelligence tools designed to identify illegal goods among legal cargo. This approach is based on the principles of machine learning and artificial intelligence, where models are trained on representative data sets to generalize patterns and make predictions.

According to research ~\cite{descalle2006analysis} the mean cargo density is expected to be about or below 0.2~g/cm$^{3}$, which means that most of the commerce goods traffic consist of low-Z materials that contain mainly C, H, O and N. Illegal materials like drugs or explosives also consist of the same elements which make the task of their detection challenging when they are hidden among the legal goods, because the Coulomb scattering of muons for these materials is small, resulting in a small deviation of the muon trajectory from a straight line.

There are several tips for shipping container loading to ensure safe transportation. A master case is a bag with bulk materials or carton box that contains multiple units, and sometimes also includes inner packs, which is another level of packaging.  To facilitate handling and storage, cargo usually placed on pallets. A pallet is a shipping platform in which multiple master cases are shipped on. The maximum weight of 20 ft container (with load) is 30,480 kg and maximum loading capacity is 28,200 kg. 

\begin{figure*}[t]
	\begin{minipage}{1.\linewidth}
		\centering
		\includegraphics[width=0.7\linewidth]{figures/abc.png}
		\vspace{-1.mm}  
	\end{minipage}		
	\begin{minipage}[t]{0.35\textwidth}
		\centering
		\includegraphics[width=1.\textwidth]{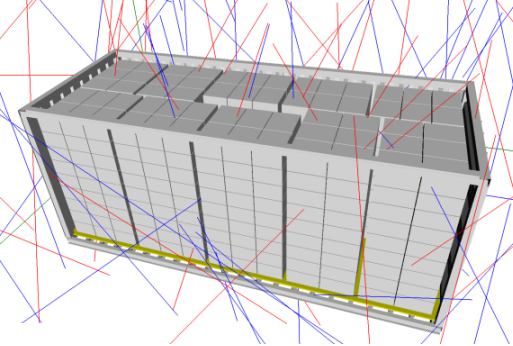}
	\end{minipage}
	\begin{minipage}[t]{0.32\textwidth}
		\centering
		\includegraphics[width=1.\textwidth]{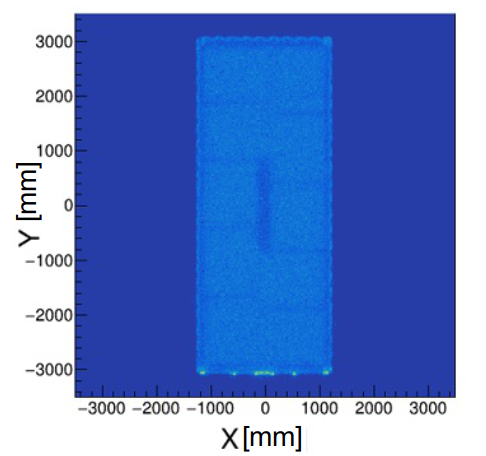}
	\end{minipage}
	\begin{minipage}[t]{0.32\textwidth}
		\includegraphics[width=1.\textwidth]{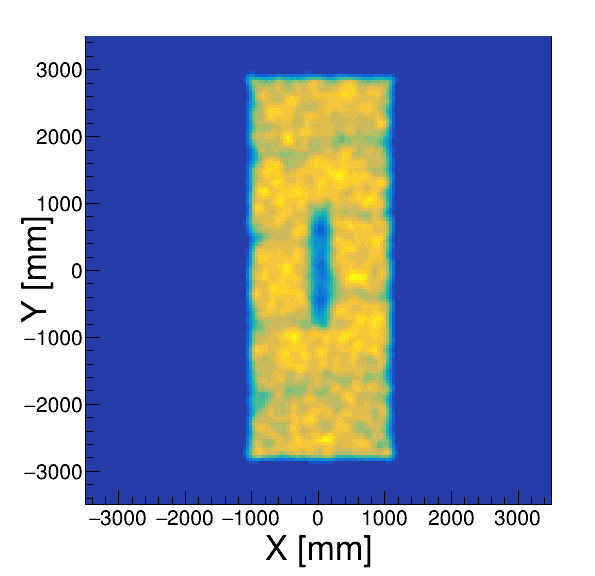}
	\end{minipage}
	\caption{(a) GEANT4 visualization of shipping container loaded cardboard boxes with cloths arranged on pallets; XY projection of PoCA scattering density of container loaded with cloths before (b) and after applying proper processing (c).}
	\label{fig:figure3}
\end{figure*}

To simulate realistic smuggling scenarios we have developed in Geant4 detailed model of shipping container ($609.6\times259.1\times243.8$ $cm^3$) loaded with cloths in cardboxes on standard pallets (b) figures~\ref{fig:figure3} (a). On Figure~\ref{fig:figure3} (b) shows 2D projection of tomographic images after applying angle cuts. The image is characterized by uniform distribution of scattering density and each palette can be resolved. 

\begin{figure*}[t]
	\begin{minipage}{1.\linewidth}
		\centering
		\includegraphics[width=0.7\linewidth]{figures/abc.png}
		\vspace{-1.mm}  
	\end{minipage}		
	\begin{minipage}[t]{0.35\textwidth}
		\centering
		\includegraphics[width=1.\textwidth]{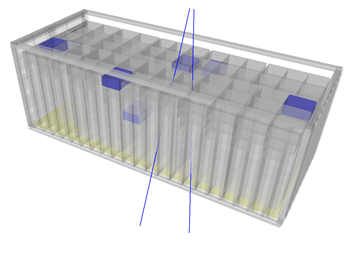}
	\end{minipage}
	\begin{minipage}[t]{0.32\textwidth}
		\centering
		\includegraphics[width=1.\textwidth]{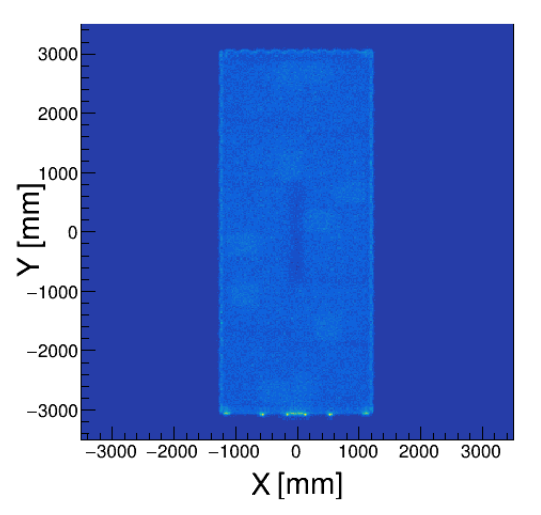}
	\end{minipage}
	\begin{minipage}[t]{0.32\textwidth}
		\includegraphics[width=1.\textwidth]{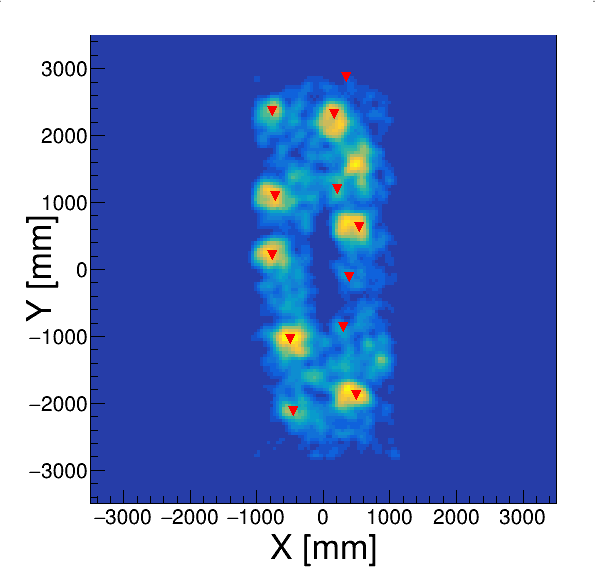}
	\end{minipage}
	\caption{(a) GEANT4 visualization of shipping container loaded cardboard boxes with cloths arranged on pallets; (b) XY projection of PoCA scattering density of container loaded with cloths;(c). GEANT4 visualization of the container in which one bag in each pallet is loaded with explosive (bags in blue color).}
	\label{fig:figure4}
\end{figure*}

We consider the scenario of illegal material hidden among boxes with legal goods. For this in each palette one randomly selected box of is loaded with explosive material (density 1.812 g/cm$^{3}$)(figure~\ref{fig:figure4}(b).

Figure ~\ref{fig:figure4}(b) shows the 2D projection of the tomographic image and some suspicious objects can be identified on this figure. After image processing to remove noise from the cargo material, objects with high scattering density are increased in contrast (figure~\ref{fig:figure4}(c)). For automatic detection of hidden explosive we have applied two-dimensional peak search functions from ROOT package which perform automatic peak searches in two-dimensional data. Most of peaks are detected. Besides real peaks of there were detected ghost peaks due to remaining noise from cargo material. 

\begin{figure*}[t]
\begin{minipage}{1.\linewidth}
	\centering
	\includegraphics[width=0.7\linewidth]{figures/abc.png}
	\vspace{-1.mm}  
\end{minipage}		
\begin{minipage}[t]{0.32\textwidth}
	\includegraphics[width=1.0\textwidth]{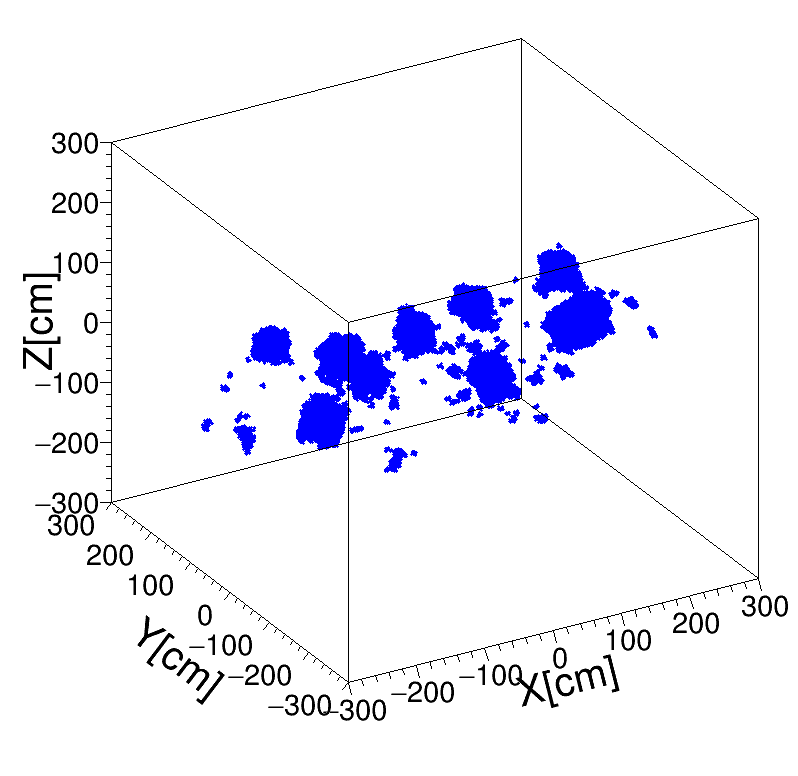}
\end{minipage}
\begin{minipage}[t]{0.32\textwidth}
	\includegraphics[width=0.9\textwidth]{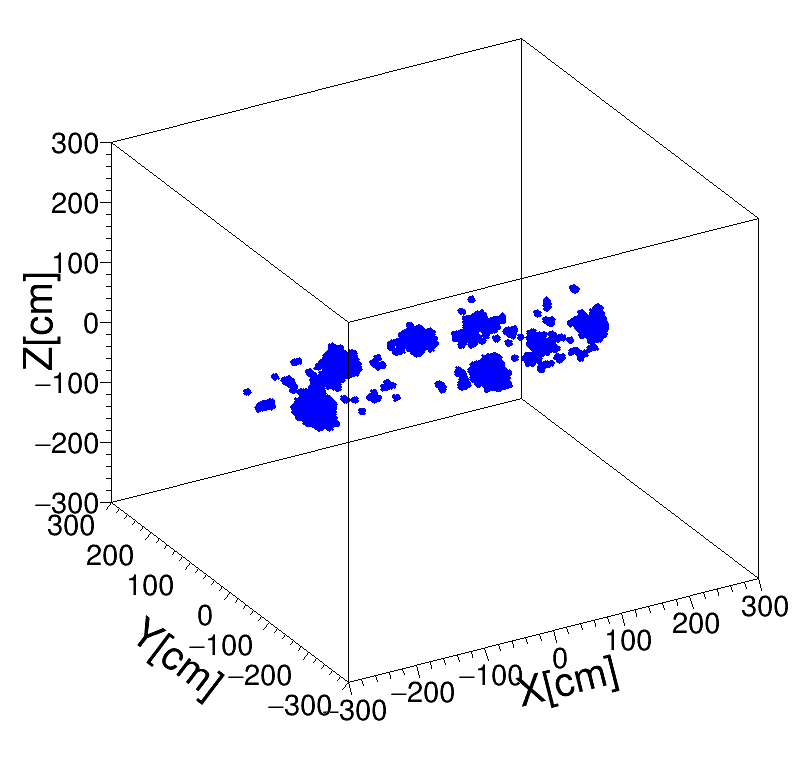}
\end{minipage}
\begin{minipage}[t]{0.32\textwidth}
	\includegraphics[width=0.9\textwidth]{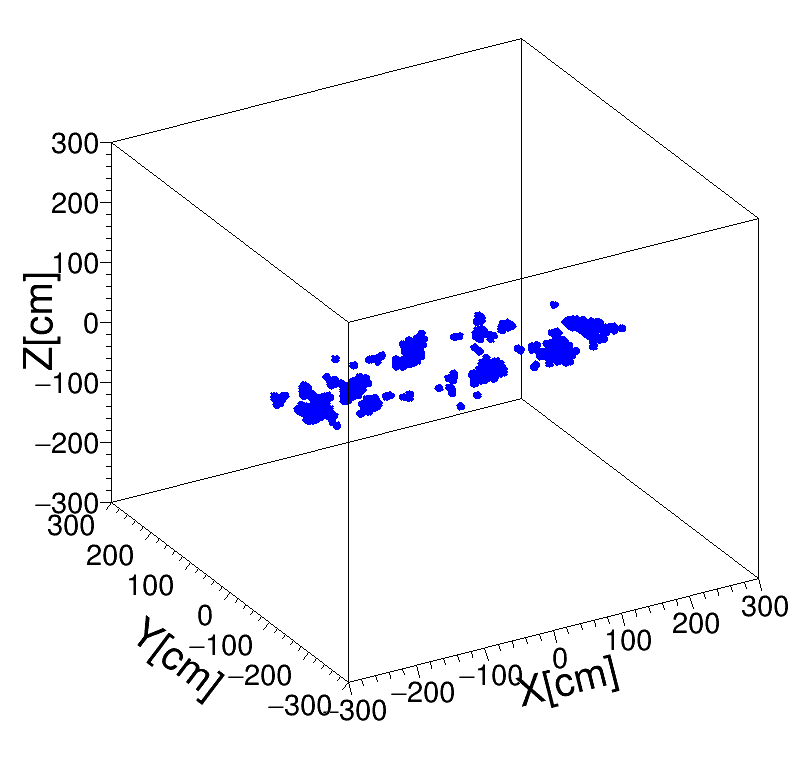}
\end{minipage}
\caption{Reconstructed muon tomographic images for scenarios of explosive hidden in container among cloths (a), bananas (b) and dry pasta (c).}
\label{fig:figure5}
\end{figure*}

\begin{figure*}
\begin{minipage}{1.\linewidth}
\centering
\includegraphics[width=0.7\linewidth]{figures/abc.png}
\vspace{-1.mm}  
\end{minipage}		
\begin{minipage}[t]{0.32\textwidth}
\includegraphics[width=1.0\textwidth]{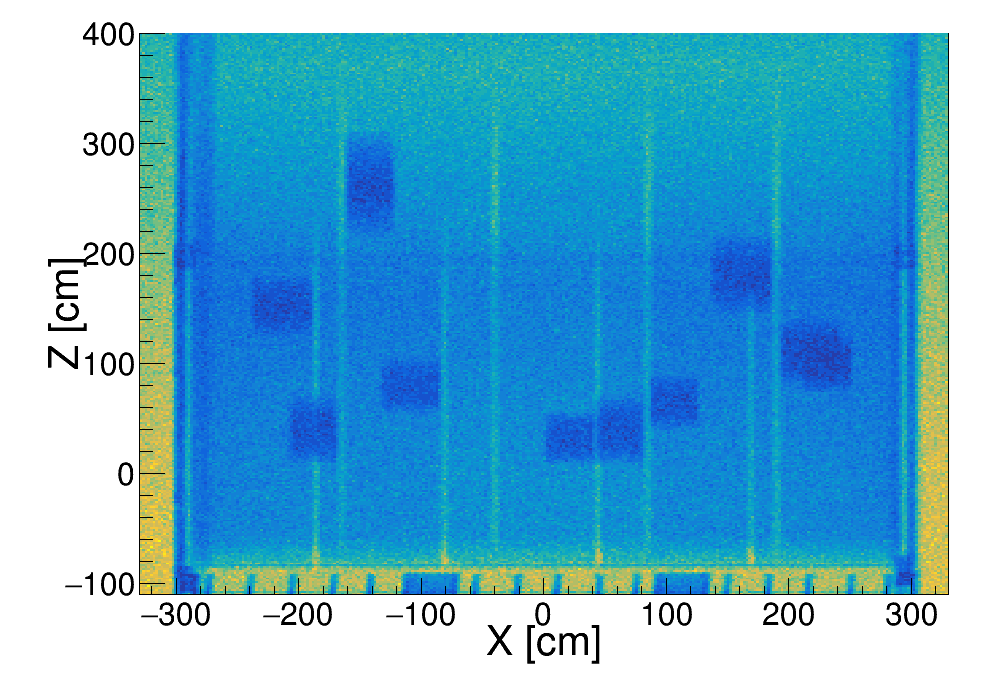}
\end{minipage}
\begin{minipage}[t]{0.32\textwidth}
\includegraphics[width=1.\textwidth]{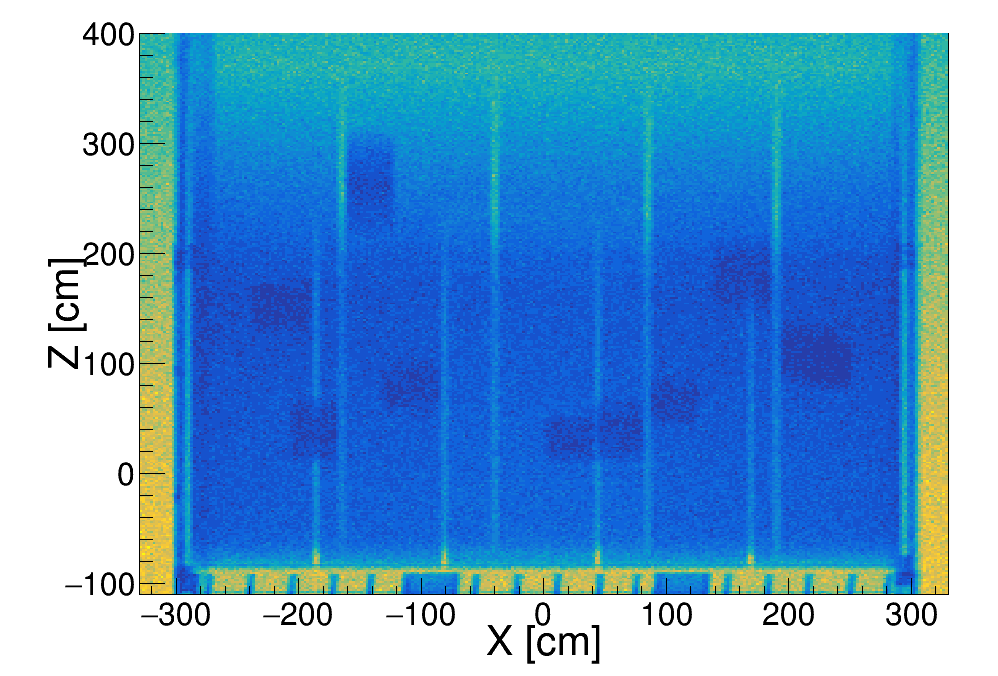}
\end{minipage}
\begin{minipage}[t]{0.32\textwidth}
\includegraphics[width=1.\textwidth]{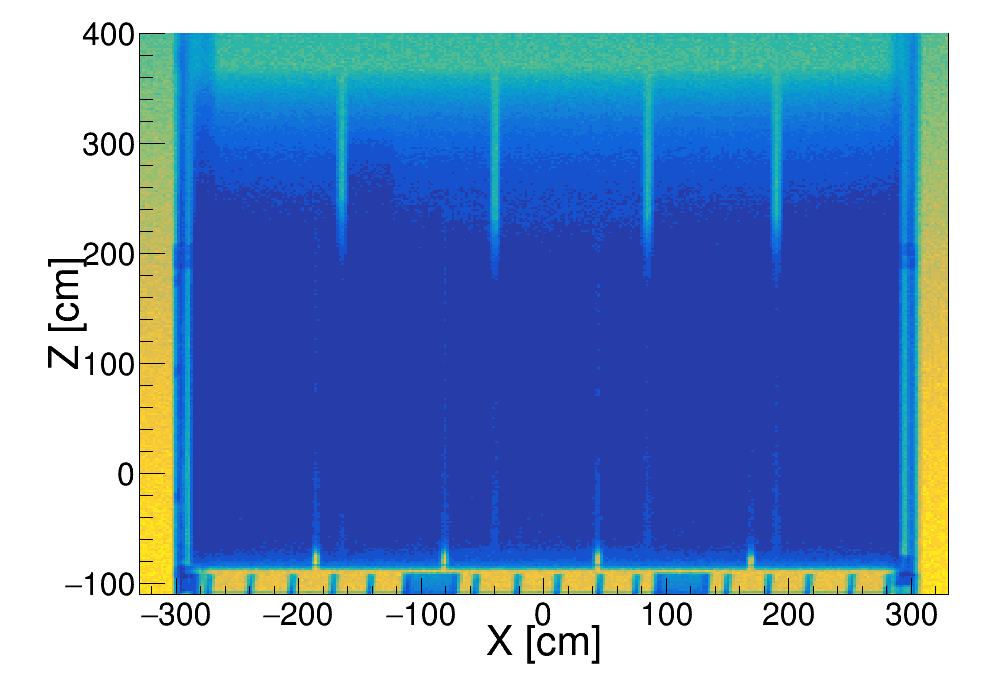}
\end{minipage}
\caption{Radiographic images simulated in Geant4 using 9 MeV X-rays for the same scenarios of explosive hidden in container among cloths (a), bananas (b) and dry pasta (c).}
\label{fig:figure6}
\end{figure*}

To test applying algorithms in different environments we simulated three scenarios: explosive hidden inside legal cargoes such as cloths, bananas and dry pasta shown in Figure~\ref{fig:figure5}(a),(b),(c), correspondingly. Reconstruction of hidden explosive is well done in cloths, worse in bananas, and unreliable for case of dry pasta. 
Figure~\ref{fig:figure6} shows for comparison  radiographic images of the same geometries simulated with Geant4 using energy spectrum of 9 MeV X-rays. As can be seen from comparing Figures~\ref{fig:figure5} and~\ref{fig:figure6} sensitivity of the two techniques is similar. 

\begin{figure*}[t]
\begin{minipage}{1.\linewidth}
\centering
\includegraphics[width=0.7\linewidth]{figures/abc.png}
\vspace{-1.mm}  
\end{minipage}		
\begin{minipage}[t]{1.\textwidth}
\centering
\includegraphics[width=0.88\textwidth]{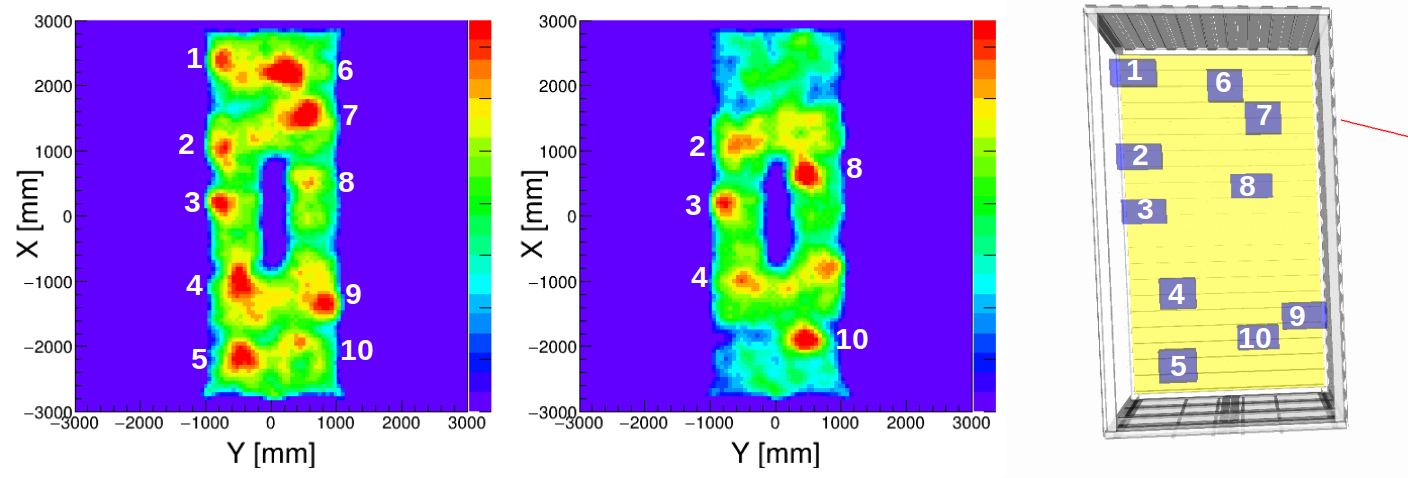}
\end{minipage}
\caption{(a) 2d projection of upper slice of tomographic image; (b) 2d projection of lower slice of tomographic image; (c) GEANT4 visualization of a transport container with the location of boxes with explosives, numbered with white numbers.}
\label{fig:figure7}
\end{figure*}

However, muon tomography has advantage on providing three-dimensional data. 
Figure~\ref{fig:figure7} shows the results of simulated scenario of explosive inside bulk dry past. Dividing the tomographic image on two slices and applying image filtering techniques all to 10 boxes (marked with white numbers) can be identified. 

\begin{figure}[ht]
\includegraphics[width=.33\linewidth]{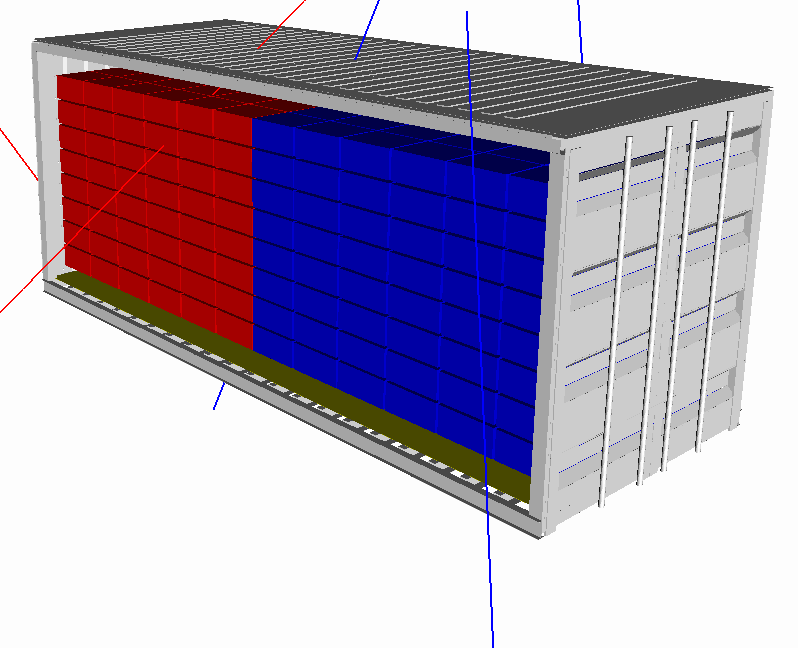}
\includegraphics[width=.33\linewidth]{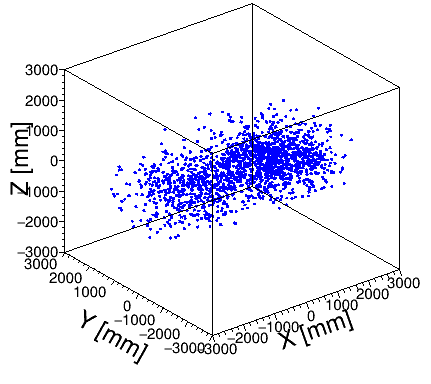}
\includegraphics[width=.33\linewidth]{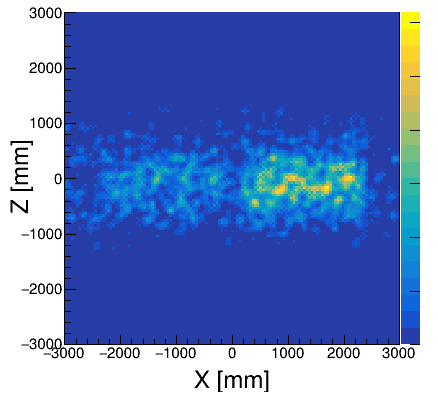}
\caption{(a) GEANT4 visualization of a shipping container half loaded with cardboard boxes of cigarettes (red) and half loaded with cardboard boxes of paper towels (blue); (b) tomographic reconstruction of a container with cigarettes and a towel; (c) 2D projection of the tomographic image after image processing.}
\label{Fig:figure8}
\end{figure}

Cigarette smuggling, also informally referred to as "buttlegging", is the illicit transportation of cigarettes or cigars from an administrative division with low taxation to a division with high taxation for sale and consumption ~\cite{wikiwebsite,global}. It often involves diversion from legal
trade channels of entire container loads, each carrying
about 10 million cigarettes. In case of such contraband cigarettes are declared as towel paper. 
To simulate a cigarette smuggling scenario, we develop a container geometry half loaded with cardboard boxes of paper towels and half loaded with boxes of cigarettes.
The bulk density of cigarettes is 
$\approx$ ~0.18 g/cm$^{3}$, the density of packaged bulk paper towels is 0.11 g/cm$^{3}$. 
The reconstructed tomographic image of this scenario is illustrated on figure~\ref{Fig:figure8}. 
It can be seen that the difference in scattering density between cigarettes and paper towels is clearly visible, allowing the detection of contraband cigarettes.

\begin{figure*}[t]
\begin{minipage}{1.\linewidth}
\centering
\includegraphics[width=0.7\linewidth]{figures/abc.png}
\vspace{-1.mm}  
\end{minipage}		
\begin{minipage}[t]{0.32\textwidth}
\centering
\includegraphics[width=1.\textwidth]{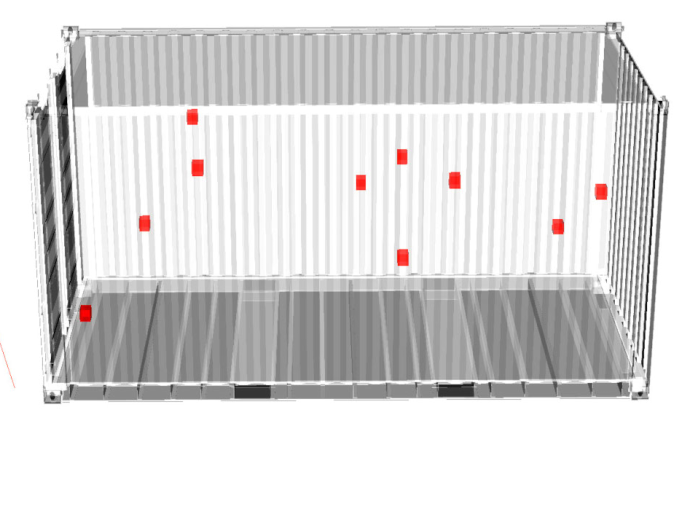}
\end{minipage}
\begin{minipage}[t]{0.32\textwidth}
\centering
\includegraphics[width=1.\textwidth]{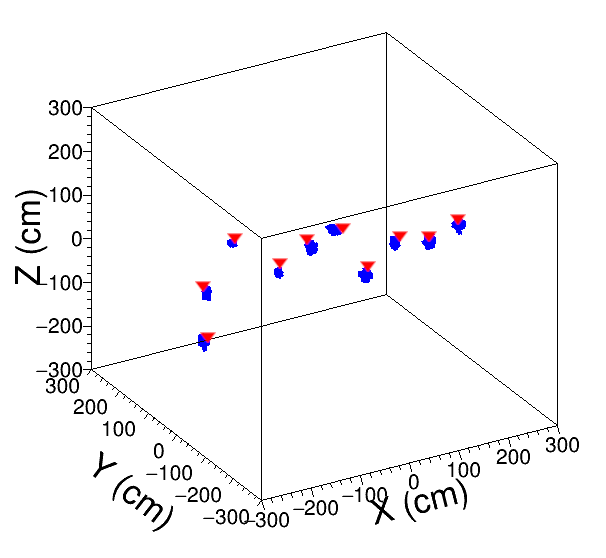}
\end{minipage}
\begin{minipage}[t]{0.32\textwidth}
\centering
\includegraphics[width=1.\textwidth]{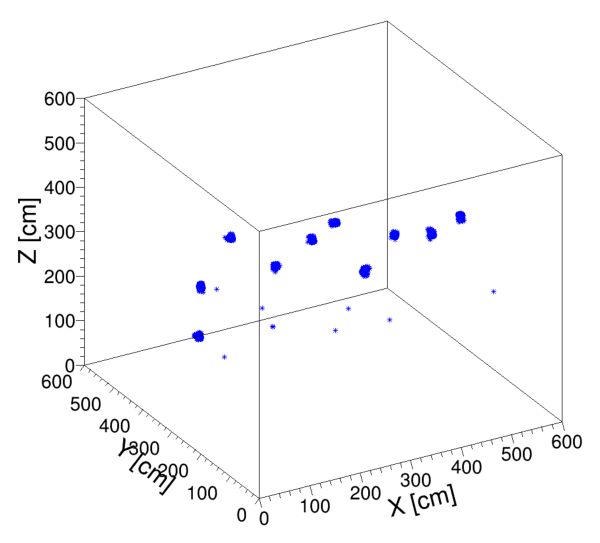}
\end{minipage}
\caption{(a) Tomographic imaging of the simulated scenario obtained with POCA of 10 cubes of SNM hidden inside of dry pasta loaded the container; (b) filtered tomographic image using adaptive threshold with detected objects using 3d peak search function. Rad triangles shows detected objects; (c) detected positions of high-Z objects in 1000 simulated data samples using 3D peak search function.}
\label{fig:figure9}
\end{figure*}

It is already have been proven that muon tomography is effective at detecting high atomic mass materials, such as Special Nuclear Materials (SNM). 
We simulated scenario of SNM concealed within a shipping container loaded with dry pasta in boxes (loading of container in weight is 27 tones). Figure~\ref{fig:figure9}(a) shows the positioning of 10 cubes of SNM, each with dimensions of 10 cm$^3$ inside boxes with dry pasta. Please note that cardboxes and container wall have been made invisible in the figure to enhance clarity. 
Results of simulation and filtered tomographic reconstruction shows (figure~\ref{fig:figure9} (b)) that all 10 cubes of SNM for this scenario can be successfully identified. Applied data processing algorithms allow fully suppress noise from container and cargo materials. For automate detection of positions of SNM cubes in the filtered tomographic image we have used the three-dimensional peak search functions from ROOT package which searches for peaks in the 3three-dimensional tomographic data. We have generated 1000 data samples for this scenario and in all samples 10 cubes were successfully detected using peak search function and their positions localized~\ref{fig:figure9}(c). Only in several cases noise signals were detected as high-Z objects.  

\section{{ Conclusions }}
\vspace{-1mm}
In this work, various scenarios of smuggling illegal low-Z and high-Z materials were investigated through Monte Carlo simulations. We tested performance of image processing tools in both cases low-Z and high-Z materials using POCA reconstruction. The tests demonstrate the validity of image processing tools and reconstruction techniques used, confirming the effectiveness of the muon tomography in combating smuggling both low- and high-Z illegal materials.
The developed algorithms are fast and suitable to be used in a real time processing of a tomographic images. The crucial aspect of developed approach is adapting threshold dependent on the scattering density observed across multiple layers around given voxel when removing the background from cargo. It implies a dynamic threshold setting that accounts for variations in scattering density in the wider environment. 
The idea of dynamically setting thresholds accounts for the variations in scattering density observed across different layers. This adaptability ensures that the algorithm remains responsive to changes in density patterns, providing a more accurate identification of anomalies.
Noise removal using the developed algorithms allows the use of peak search function from the ROOT data analysis package to automatically detect anomalies in tomographic images caused by hidden objects. Automatic anomaly detection can then trigger an alarm and aims to improve the detection and localization of hidden objects in tomographic data.
In this work we have demonstrated that muon scattering tomography is a promising technique for combating smuggling using shipping containers.

\section{Acknowledgements}
The authors acknowledge partial funding from the EU Horizon 2020 Research and Innovation Programme under grant agreement no. 101021812 (“SilentBorder”).

\bibliographystyle{JHEP}
\bibliography{bibliography}
\end{document}